\documentclass[journal,10pt]{IEEEtran}
\usepackage{amsfonts}
\IEEEoverridecommandlockouts

\ifCLASSINFOpdf
\else
\fi
\usepackage{epsfig}
\usepackage{graphicx}
\usepackage{psfig}
\usepackage{epsf}
\usepackage[cmex10]{amsmath}
\usepackage{booktabs}
\usepackage{fancyhdr}
\usepackage{booktabs}
\usepackage{stfloats}
\usepackage{algorithmic}
\usepackage[ruled]{algorithm2e}
\usepackage{graphicx}
\usepackage{subfigure}
\usepackage{color}
\newtheorem{myDe f}{Definition} 

\newtheorem{remark}{Remark}
\makeatletter

\newcommand{\Rmnum}[1]{\expandafter\@slowromancap\romannumeral #1@}
\makeatother

\begin{document}
 \title{Subcarrier Assignment Schemes Based on Q-Learning in Wideband Cognitive Radio Networks}
\author{\IEEEauthorblockN{Yuan Zhou, Fuhui Zhou, \emph{Member, IEEE}, Yongpeng Wu, \emph{Senior Member, IEEE}, Rose Qingyang Hu, \emph{Senior Member, IEEE}, and Yuhao Wang, \emph{Senior Member, IEEE}}
\thanks{ Copyright (c) 2015 IEEE. Personal use of this material is permitted. However, permission to use this material for any other purposes must be obtained from the IEEE by sending a request to pubs-permissions@ieee.org.

Manuscript received July. 11, 2019; revised Sept. 27, 2019 and accepted Nov. 12, 2019. The associate editor coordinating the review of this paper and approving it for publication was Bomin Mao. (The Corresponding author is Fuhui Zhou.)

F. Zhou is with the College of Electronic and Information Engineering,
Nanjing University of Aeronautics and Astronautics, Nanjing, 210000, P. R.
China. (e-mail: zhoufuhui@ieee.org). Y. Zhou and Y. Wang are with Nanchang University, P. R. China. (e-mail: 6106216037@email.ncu.edu.cn, wangyuhao@ncu.edu.cn). R. Hu is with Utah State University, USA. (e-mail: rose.hu@usu.edu).  Y. Wu is with Shanghai Key Laboratory of Navigation and Location Based Services, Shanghai Jiao Tong University, Minhang, 200240, China. Y. Wu is also with State Key Laboratory of Integrated Services Networks,
Xidian University, Xian, China.(Email: yongpeng.wu@sjtu.edu.cn).}

\thanks{This work was supported in part by the National Natural Science Foundation of China ( 61701214, 61701301 and 61661028), the Intel Corporation, the US National Science Foundation under Grant EARS¨C1547312, the Excellent Youth Foundation of Jiangxi Province under Grant 2018ACB21012, and the open research project of State Key Laboratory of Integrated
Services Networks (Xidian University) under Grant ISN20-03. }}

\maketitle
\begin{abstract}
Subcarrier assignment is of crucial importance in wideband cognitive radio (CR) networks. In order to tackle the challenge that the traditional optimization-based methods are inappropriate in the dynamic spectrum access environment, an independent Q-learning-based scheme is proposed for the case that the secondary users (SUs) cannot exchange information while a collaborative Q-learning-based scheme is proposed for the case that information can be exchange among SUs. Simulation results show that the performance achieved with the proposed collaborative Q-learning-based assignment is better than that obtained with the proposed independent Q-learning-based assignment at the cost of the computation cost.

\end{abstract}
\begin{IEEEkeywords}
Wideband cognitive radio, subcarrier assignment, Q-learning.
\end{IEEEkeywords}
\IEEEpeerreviewmaketitle
\section{Introduction}

Spectrum scarcity is increasingly severe due to the emergence of diverse wideband services and the proliferation of mobile devices \cite{Elias Z. Tragos}, \cite{N. Kato}. In order to address this issue, cognitive radio (CR) has been received significant attention from both academia and industry \cite{Z. M. Fadlullah}. It enables the secondary users (SUs) to coexist with the primary users (PUs) in the same spectrum band of the PUs as long as the interference level caused to the PUs is smaller than that of the tolerable threshold of the PUs. In this case, in order to protect the quality of service (QoS) of the PUs, the transmit power of the SUs cannot be too large, which limits the performance of the SUs. As a promising network, wideband cognitive radio network (CRN) that allow SUs to use multiple subcarrier bands is efficient to enhance the performance of SUs \cite{Elias Z. Tragos}, \cite{N. Kato}.

In CRNs, subcarrier assignment is of crucial importance. It not only can efficiently protect the PUs from harmful interference, but also can improve the performance of the SUs \cite{Elias Z. Tragos}. Up to now, most of the existing works designed subcarrier assignment schemes based on optimization methods. Specifically, the subcarrier assignment schemes were obtained by formulating and solving subcarrier assignment problems \cite{Fuhui Zhou}-\cite{H. Zhang}. For example, in \cite{Fuhui Zhou} and \cite{R. Fan}, subcarrier assignment in wideband CR was formulated as a mixed-integer programming problem. In \cite{Fuhui Zhou}, the channel allocation problem was transformed into a convex optimization problem by relaxing subchannel sharing constraints. In \cite{R. Fan}, the authors exhaustively searched all the combinations of assignment to find the optimal solution.  The channel allocation problem for heterogeneous CR was studied in \cite{H. Zhang}, and was solved in a similar way as that way in \cite{Fuhui Zhou}. However, on one hand, the increase of the number of subchannels and that of SUs result in the significant increase of the algorithm complexity of the optimization-based methods. On the other hand, those works were based on specific mathematical models, which cannot match the practical scenarios. In this case, those optimization-based subcarrier assignment schemes cannot achieve a sound performance and even cannot work in practice.

In order to tackle the above-mentioned problems, the method based on machine learning tools have became a primary option. Deep learning has been used to configure and manage networks in a intelligent way \cite{B. Mao}, \cite{Fadlullah}. Recently, Deep learning have been applied in CRNs and other networks for realizing intelligent network managements  \cite{Z. M. Fadlullah}, including traffic control \cite{B. Mao} and routing design \cite{Fadlullah}. Motivated by the fact that reinforcement learning (RL) is powerful and efficient to make decisions, RL has been used to design subcarrier assignment schemes. As a simple model-free RL algorithm, Q-learning is a popular solution for decision problems when the number of states and actions are both finite. Since it dose not need to model the environment, Q-learning has been used in conventional networks to solve the resource allocation problems, including power control and subcarrier assignment problems. In \cite{R. Amiri}, Q-learning has been applied for power optimization in heterogeneous wireless networks while the authors of \cite{L. Xiao} have proposed the power scheme based on Q-learning in dynamic NOMA transmission game and used Dyna architecture and hotbooting techniques to achieve a faster learning speed. 
To solve the channel assignment problem, in \cite{L. M. Bello} and \cite{N. Morozs}, schemes based on Q-learning for spectrum access in cellular networks were proposed.


Up to now, to the authors' best knowledge, subcarrier assignment in CRNs has not been proposed based on RL. Moreover, due to the inefficient exploration, there may exist the case that multiple SUs contend for the same subcarrier at the same frame. It can reduce the performance of SUs. In order to overcome it, collaborative RL is promising \cite{Oshri Naparstek}.

In this letter, two distributed subcarrier assignment schemes are proposed based on Q-learning in wideband CRNs. Specifically, a distributed independent Q-learning-based scheme is designed for the case that SUs cannot exchange information. Moreover, in order to improve the efficiency of exploration, a collaborative Q-learning scheme is proposed when SUs can exchange information. Simulation results show that the convergence of the collaborative Q-learning is greater than that of the independent Q-learning when the number of SUs is greater than that of the subchannels.

The rest of this paper is organized as follows. Section \uppercase\expandafter{\romannumeral2} presents the system model. In Section \uppercase\expandafter{\romannumeral3}, two  assignment Q-learning schemes are proposed. Section \uppercase\expandafter{\romannumeral4} presents simulation results and this letter concludes with Section \uppercase\expandafter{\romannumeral5}.

\section{System Model}
A wideband CRN is considered, where the primary network coexists with the secondary network under the spectrum sharing paradigm. In the secondary network, there is one cognitive base station (CBS) and $K$ SUs while the  primary network consists of $N$ PUs and one primary base station (PBS). As shown in Fig. 1, in order to serve $N$ PUs and enhance the performance of SUs, $M$ non-overlapping orthogonal subchannels are considered and each PU occupies one subchannel in order to avoid mutual interference among PUs. It is assumed that the mutual interference between PUs and SUs is intolerable once two or more SUs access the same subchannel.

The transmission rate of the $k$th SU which transmits over the $m$th channel is given as
\begin{align}
&{C_{m,k}}={\frac{B}{N}}{\mathbf{1}\left(m\right)}{\log_2 \left(1+\frac{P_{m,k}g_{m,k}} {P^{PU}_{m}z_{m,k}+ \sigma_{k}^2}\right)},
\end{align}
where $B$ and $g_{m,k}$ denote the total bandwidth of the primary network and the channel power gain between the CBS and the $k$th SU over the $m$th channel, respectively. Let $P_{m,k}, m\in\mathcal{M}, k\in\mathcal{K}$ denote the transmit power of the $k$th SU over the $m$th channel, where $\mathcal{K}=\{1,2,\dots,K\}$ denotes the set of SUs and $\mathcal{M}=\{1,2,\dots,M\}$ denotes the set of subchannels. $z_{m,k}\text{ and }\sigma_{k}^2$ denote the channel power gain between the PBS and the $k$th SU over the $m$th channel and the variance of additive white Gaussian noise at the $k$ SU, respectively. $\mathbf{1}(m):\mathcal{M}\to\{0,1\}$ is the indicator function, given as
\begin{align}
{\mathbf{1}(m)} =\left\{\begin{array}{rcl} {1,} && {n_{m} = 1} \\ {0,} && { \text{otherwise}}\end{array}\right.
\end{align}
where $n_{m}$ denotes the number of the SUs that access the $m$th channel.

When the $k$th SU chooses the $m$th channel, the throughput of the CRN can be given as
\begin{align}
&{\gamma_{total}}={\displaystyle\sum_{m=1}^{M}\sum_{k=1}^{K}C_{m,k}}.
\end{align}

\begin{figure}[!t]
\centering
\includegraphics[width=2.0 in]{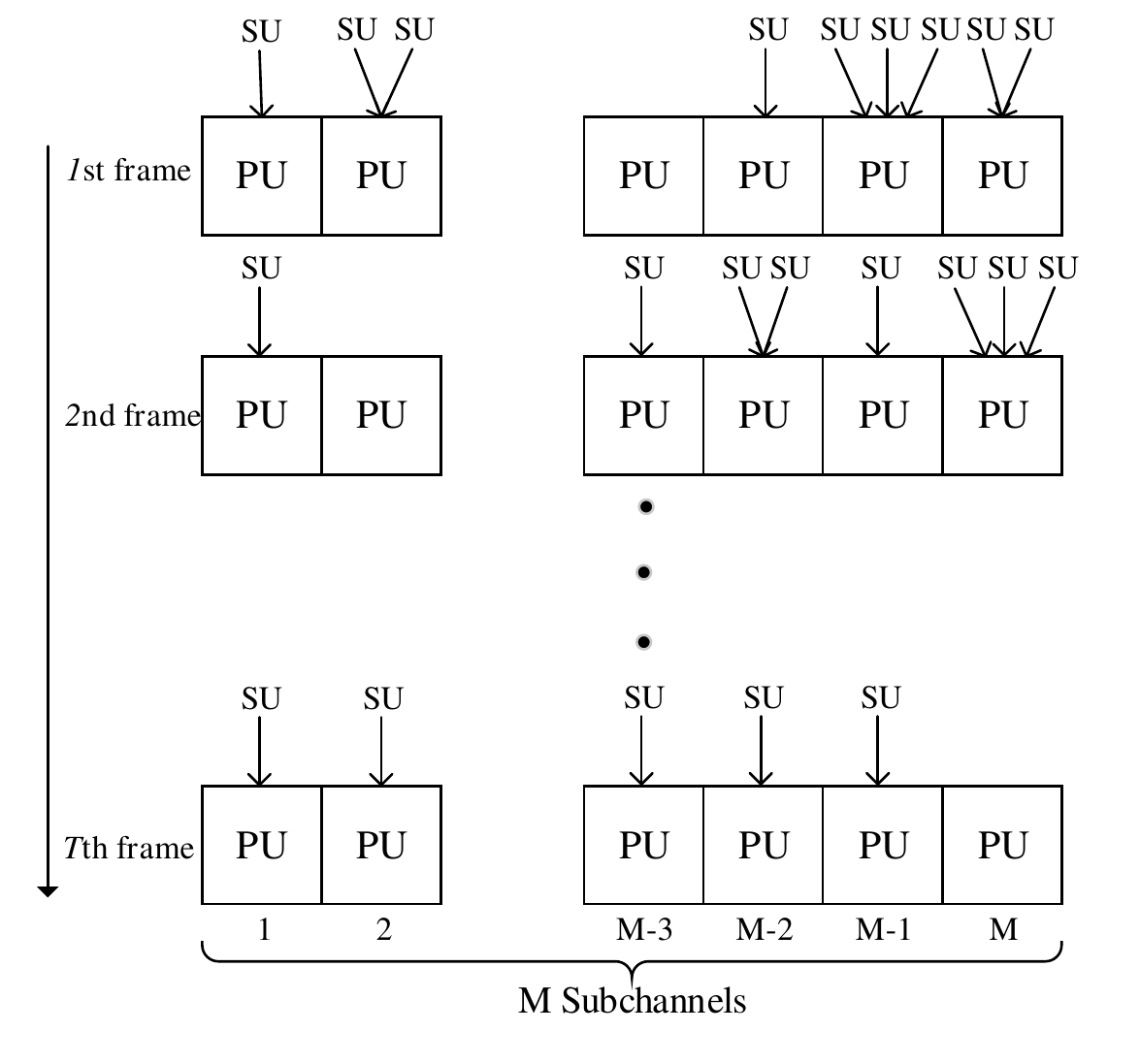}
\caption{An example of subcarrier allocation in wideband CRNs.}
\label{fig.1}
\end{figure}

\section{Q-learning-Based Subcarrier Assignment}
\subsection{Q-Learning}

RL enables an agent to learn by interacting with the environment \cite{Lucian Busoniu}. The control process in RL usually can be modeled as the Markov decision process (MDP). It is formally described as a tuple $(S, A, P, R)$, where $S,A,P \text{ and } R$ respectively denote the finite set of states, the finite set of actions, the transition probabilities set and the reward function \cite{Lucian Busoniu}.

As an off-policy RL method, Q-learning employs the behavior policy $\pi$ (e.g., greedy-$\epsilon$) to generate the samples for learning and updating another policy $\mu$ (e.g., greedy). The goal of the agent exploiting Q-learning is to maximize the expected discounted reward at each time step $t$ \cite{Lucian Busoniu}, given as
\begin{align}
&{R_{t}}={\mathbb{E}\{\displaystyle\sum_{j=0}^{\infty}\gamma^{j} r_{t+j+1}\}},
\end{align}
where $\gamma\in[0,1)$ is the discount factor and $\mathbb{E}\{\cdot\}$ represents the expected value. In this letter, the discount factor $\gamma$ is equal to zero. Once the state-action pair and the policy $\pi$ are given, the Q-value function can be given as
\begin{align}
&{Q^{\pi}(s_t,a_t)}={\mathbb{E}\{\displaystyle\sum_{j=0}^{\infty}\gamma^{j} r_{t+j+1}|s_{t}=s,a_{t}=a,\pi \}}.
\end{align}
Let $Q^{\pi^{\ast}}(s_t,a_t)$ denote the Q-value following the optimal policy. The optimal policy $\pi^{\ast}$ can be derived as $\pi^{\ast}=arg\max_{a}Q^{\pi^{\ast}}(s_{t},a_{t})$, where $arg\max$ represents that the function is maximized. Q-learning evaluates the value by using a greedy policy, given as
\begin{align}
&{Q_{t}}={Q_{t}}+{\alpha[r_{t}+\gamma\displaystyle\max_{a_{t+1}} Q_{t+1} - Q_{t}]},
\end{align}
where $\alpha\in(0,1]$ is the learning rate.

The state of the $k$th SU at the $t$th frame $a_t\in A=\mathcal{M}$ can be defined as the channel that it decides to occupy. The state of the $k$th SU in the $t$th frame $s_{t}\in S=\mathcal{M}$ can be defined as the channel that the SU attempts to occupy at this frame, i.e. $s_t=a_{t-1}$. From the definition of the state and action, it is obvious that the state at this frame is decided by the action of the previous frame and the discount factor $\gamma$ is set to zero. Thus, Q values are completely determined by immediate rewards. The immediate rewards can be influenced by the system state, the SU's state and the SU's action. However, the SU's state influences the subcarrier access for the next frame only when multiple SUs contend for the same subcarriers and these SUs contend for another subcarrier once again. In wideband CRNs, the number of subcarrier is large. Therefore, the probability that a SU contends with another SU for the same subcarrier twice is quite low. In order to reduce the cost of computing and space, the reward for the SUs that have different states and same action are considered the same.

\vbox{}

\subsection{Independent Q-Learning-Based Subcarrier Assignment}

In the multi-user scenarios, the spectrum access problem can be formulated as a Markov game, which can be expressed as $\mathcal{G}=(\mathcal{S}, \mathcal{A}_{1}, \cdots, \mathcal{A}_{K}, f, r_{1}, \cdots, r_{K})$. $\mathcal{S}$ is a finite set of system states $s_s$, $\mathcal{A}_{k},k\in\mathcal{K}$ denotes the action set of the $k$th agent, $f$ and $r_{k}$ represent the state transition probability function and the reward functions of the $k$th agent, respectively \cite{Lucian Busoniu}. The state of CR $s^i\in\mathcal{S}$ is defined as $s^i=\mathbf{a}^i$, where $\mathbf{a}^i=(a_1,a_2,\dots, a_k)$ denotes a joint action profile. $a_k\in\mathcal{K}$ denotes the action of the $k$th SU, and the set of the action profiles is $\mathbf{A}=\otimes\mathbf{A_k}$, where $\otimes$ is the Cartesian product.

In order to protect PUs from intolerable interference and improve the performance of SUs, considering that SUs cannot exchange their information due to the limit resource, each SU exploits independent Q-learning to maximize itself expected discounted reward. At the beginning of the game, the Q-values of all SUs are initialized to zero and each SU randomly selects an action $a_k$ and executes it, The reward function is 
\begin{align}
&{r_{i,k}(s_{t},a_{t})} =\left\{\begin{array}{rcl} {+1,} && {\text{if transmission succeeds},} \\ {-1,} && {\text{otherwise}.}\end{array}\right.
\end{align}

After updating the Q-values for the first time using (6) where $\gamma=0$, each SU selects an action by using the policy $\pi_{in}$ at the subsequent frame $t$. The policy $\pi_{in}(s_t)$ is
\begin{align}
\setcounter{equation}{7}
{\pi_{in}(s_{t})} = \left\{ \begin{array}{l}
 \text{arg}\max_{a_t}Q(s_t, a_t),\text{  }\text{ }\text{ } {\exists \text{Q}(s_t,a_t) > 0,}\\
\pi_{rand}(s_{t}), \text{  }\text{ }\text{ }\text{  }\text{ }\text{ }\text{  }\text{ }\text{ }\text{ }\text{ }\text{ }\text{  }\text{ }\text{ }{\text{otherwise.}}
\end{array} \right.
\end{align}
where $\pi_{rand}(s_{t})$ denotes
that an action is randomly selected and $\exists Q(s_t,a_t)>0$ represents that there exists a Q-value, which is greater than zero.

\vbox{}

\subsection{Distributed Collaborative Q-Learning-Based Subcarrier Assignment}

For the independent Q-learning scheme, the efficiency of exploration is low, which can result in the case that multiple SUs contend for the same subcarrier. In order to tackle this issue, a distributed collaborative Q-learning-based subcarrier assignment scheme is proposed. There are several different characters of collaboration, in this letter, the collaboration is reflected in the message exchange among SUs.

For the collaborative Q-learning subcarrier assignment scheme, the SU performs the same initial operation as that of the independent Q-learning subcarrier assignment scheme. To reduce the signaling overhead and complexity costs, the information exchange interval $\Delta$ and the information exchange frame $t_{ex}$ are defined. The $t_0$th frame is defined as $t=t_0$. And the frame that the information is exchanged is defined as the information exchange frame, which satisfies $t_{ex}\text{ }mod\text{ }(\Delta+1)=0 \text{ and } t_{ex}\neq 0$, where $mod$ represents the modulo operation that finds the remainder after division of one number by another. So there are $\Delta$ frames between the adjacent information exchange frames. Other frames that the information is not exchanged (i.e., $t\text{ }mod\text{ }(\Delta+1)\neq0$) is defined as the general frame. For example, if $\Delta=2$ and the current frame is the second frame, and $t=2$. The remainder of $2$ divided by $\Delta+1$ is $2$, which is not equal zero. Thus, this frame is a general frame that the information is not exchanged. At the very beginning of the subcarrier assignment, each SU has a Q-table in which the Q-values are initialized to zero. Then each SU randomly selects a channel. If the frame $t$ is an information exchange frame, before the beginning of the next frame, the SU can obtain all channels' states of the previous frame according to an extra channel, where the channel state is transmitted by the CBS and the channel state has $M$ bits message defined as $\textbf{CS} = [\text{CS}_1,\text{CS}_2,\dots,\text{CS}_M]$. If the $k$th SU successfully accesses the $m$th channel at the previous frame, the $m$th bit can be set to one. Otherwise, the $m$th bit is zero. At the information exchange frame, for the SU that all Q-values are less or equal to zero, the reward function is
\begin{align}
&{r_{j}(s_t, a_t)} =\left\{\begin{array}{rcl} {+1,} && {\text{CS}_m=1\text{ and }j=m} , \\  {-1,} && {\text{CS}_j=1\text{ and }j\neq m} , \\{0,} && {\text{otherwise}} \end{array}\right.
\end{align}
When the $k$th SU selects the $m$th channel and $\text{CS}_m=1$, the $k$th SU only needs to update its one Q-value $Q_{t}(m)$ by
\begin{align}
&{Q_{t}(m)}={Q_{t}(m)}+{\alpha[r_{t,m}- Q_{t}(m)]}
\end{align}
where $r_{t,m}=r_{m}(s_t,a_t)$. When the $k$th SU that selects the $m$th channel and the $\text{CS}_m=0$, the $k$th SU should update its all Q-values according to (9) at an information exchange frame. At an general frame,
the update method of SUs is the same as that in the independent Q-learning method. The difference is the reward, given by
\begin{align}
&{r_{m,k}(s_{t},a_{t})} =\left\{\begin{array}{rcl} {+1,} && {\text{CS}_m=1,} \\ {0,} && {\text{otherwise}.}\end{array}\right.
\end{align}
Note that for SUs that success in accessing the channel, there is no need to update the Q-values. Then the SU selects an action by using policy $\pi_{cq}$ at the subsequent frame $t$. The policy $\pi_{cq}$ can be given as
\begin{align}
\setcounter{equation}{11}
{\pi_{cq}(s_{t})} = \left\{ \begin{array}{l}
 \text{arg}\max_{a_t}Q(s_t, a_t),\text{  }\text{ }\text{ } {\exists \text{Q}(s_t,a_t) > 0},\\
\pi_{eg}(s_{t}), \text{  }\text{ }\text{ }\text{  }\text{ }\text{ }\text{  }\text{ }\text{ }\text{ }\text{ }\text{ }\text{  }\text{ }\text{ }\text{ }\text{ }\text{  }{\text{otherwise}}.
\end{array} \right.
\end{align}
where the $\epsilon-greedy$ policy is adjusted to adapt the access channel problem, given as
\begin{align}
\setcounter{equation}{12}
{\pi_{eg}(s_{t})} = \left\{ \begin{array}{l}
 \text{arg}\max_{a_t}Q(s_t, a_t),\text{  }\text{ }\text{ } {x<1-\epsilon},\\
\pi_{R}(s_t), \text{  }\text{ }\text{ }\text{  }\text{ }\text{ }\text{  }\text{ }\text{ }\text{ }\text{ }\text{ }\text{  }\text{ }\text{ }\text{ }\text{ }\text{  }{x>\epsilon}.
\end{array} \right.
\end{align}
where $\pi_{R}(s_t)$ is the policy that randomly selects an action $a_{t},a_{t}\in\mathcal{Q}^{-}$, $\mathcal{Q}^{-}=\mathcal{A}-\mathcal{Q}^{+}$ and $\mathcal{Q}^{+}$ is the set of the actions that the Q-value is greater or equal to zero. And the random variable $x\sim U(0,1)$ is used to ensure an action that is randomly selected with a probability $\epsilon$. The policy for generating the trajectories in the standard Q-learning is different from the policy $\pi_{in}$. The reason is that when the SU succeeds to access the $m$th channel, the SU can occupy the $m$th channel in the subsequent frames. Therefore, for the SU that has a Q-value that is greater than zero, exploration is unnecessary and inefficiency.

\begin{remark}
Due to the introduction of multiple agents to the MDP model, not only the SU's state transfers and the action that the SU selects, but also the joint action of all agents determines the reward and changes in the environment. In order to tackle this complicated issue, we use the Markov games. In MDPs, each agent tries to maximize its expected discounted reward while in Markov games, the reward criteria also depends on the policies of other agents \cite{Vrancx P.}. In the independent Q-learning scheme, each SU updates its Q-values independently while the rewards in the collaborative scheme not only depends on the previous state, current state and action, but also are influenced by other SUs' actions.
\end{remark}

\begin{remark}

In the practical system, our proposed independent Q-learning method can work in the case that there is no resource for SUs to exchange the information. The SUs update their access strategies based on their own reward. For our proposed collaborative Q-learning scheme, it can work in the case that SUs can exchange information among each other. They update their decisions and rewards based on the exchanged information. Those two algorithms can be implemented in the software defined radio (SDR) platform \cite{S. K. Sharma}.

In \cite{J. Mitola}, the authors identified SDR as the ideal environment for the deployment of CRs. Depending on specific requirements of deployment, there are several software and hardware platforms based on SDR for the implement of CRs. With those highly flexible platforms, the CRN can be constructed through programming. For example, GNU Radio is an open source SDR platform that provides a complete development environment to create radio systems. Our proposed algorithms can be implemented in this platform with C++ and Python.
\end{remark}

\begin{figure}[!t]
\centering
\includegraphics[width=2.3 in]{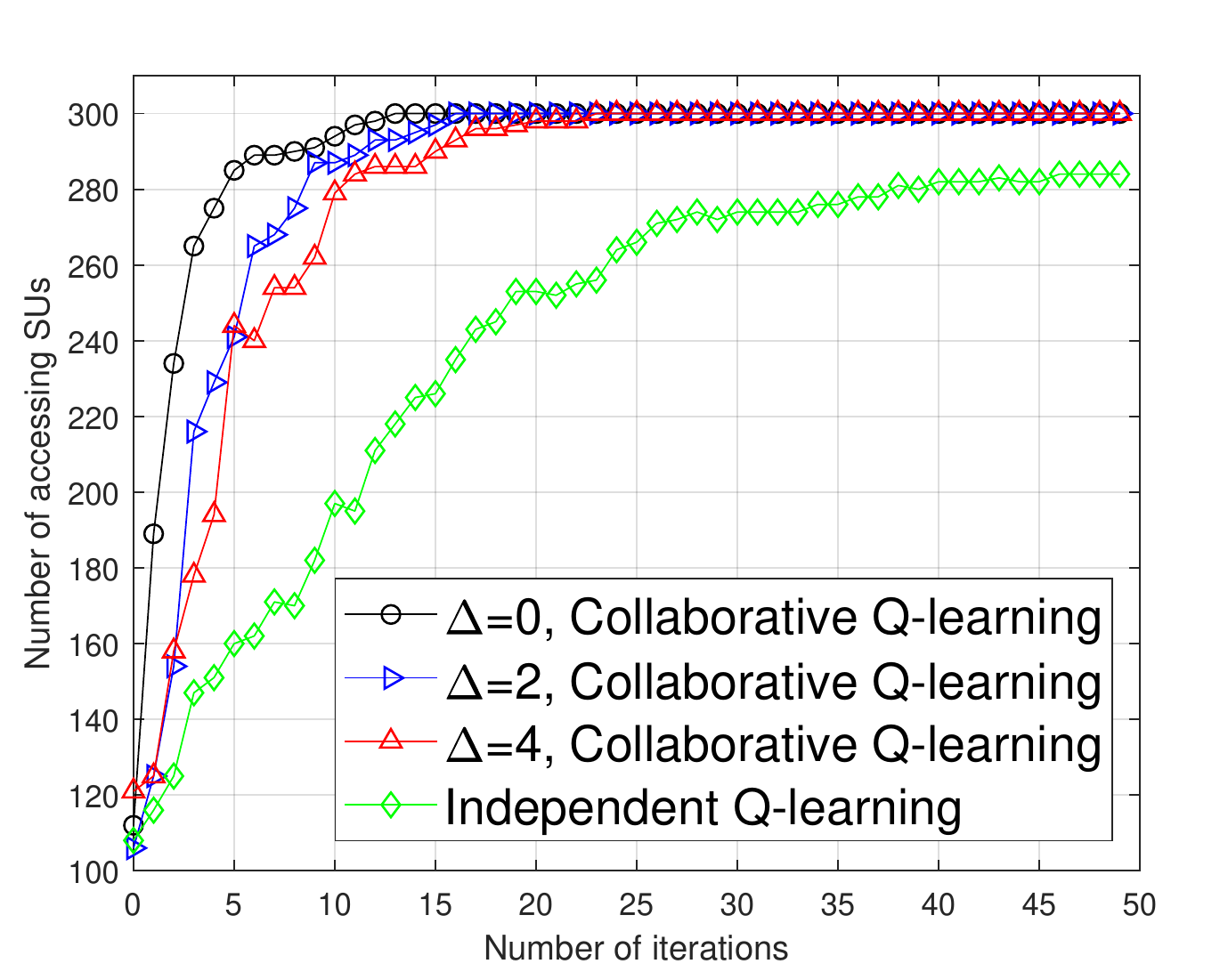}
\caption{The number of accessing SUs versus the number of iterations with the independent Q-learning and collaborative Q-learning-based schemes ($M = K = 300$, $\text{iterations} = 50$).}
\label{fig.1}
\end{figure}

\begin{figure}[!t]
\centering
\includegraphics[width=2.2 in]{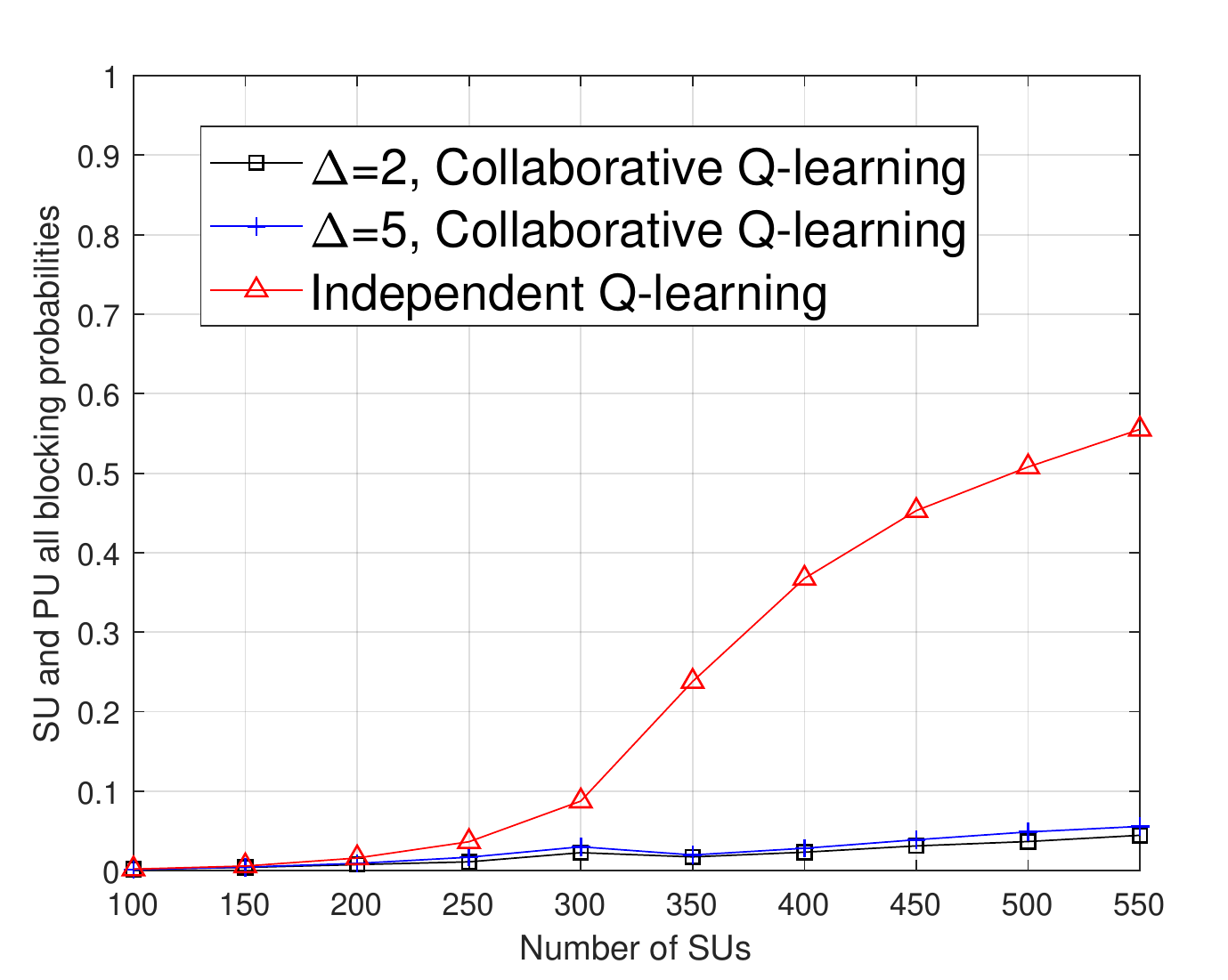}
\caption{The probabilities of SU and PU all blocking versus the numbers of SUs with the independent Q-learning and collaborative Q-learning-based schemes ($M = 300$, $\text{iterations} = 50$).}
\label{fig.1}
\end{figure}

\begin{figure}[!t]
\centering
\includegraphics[width=2.2 in]{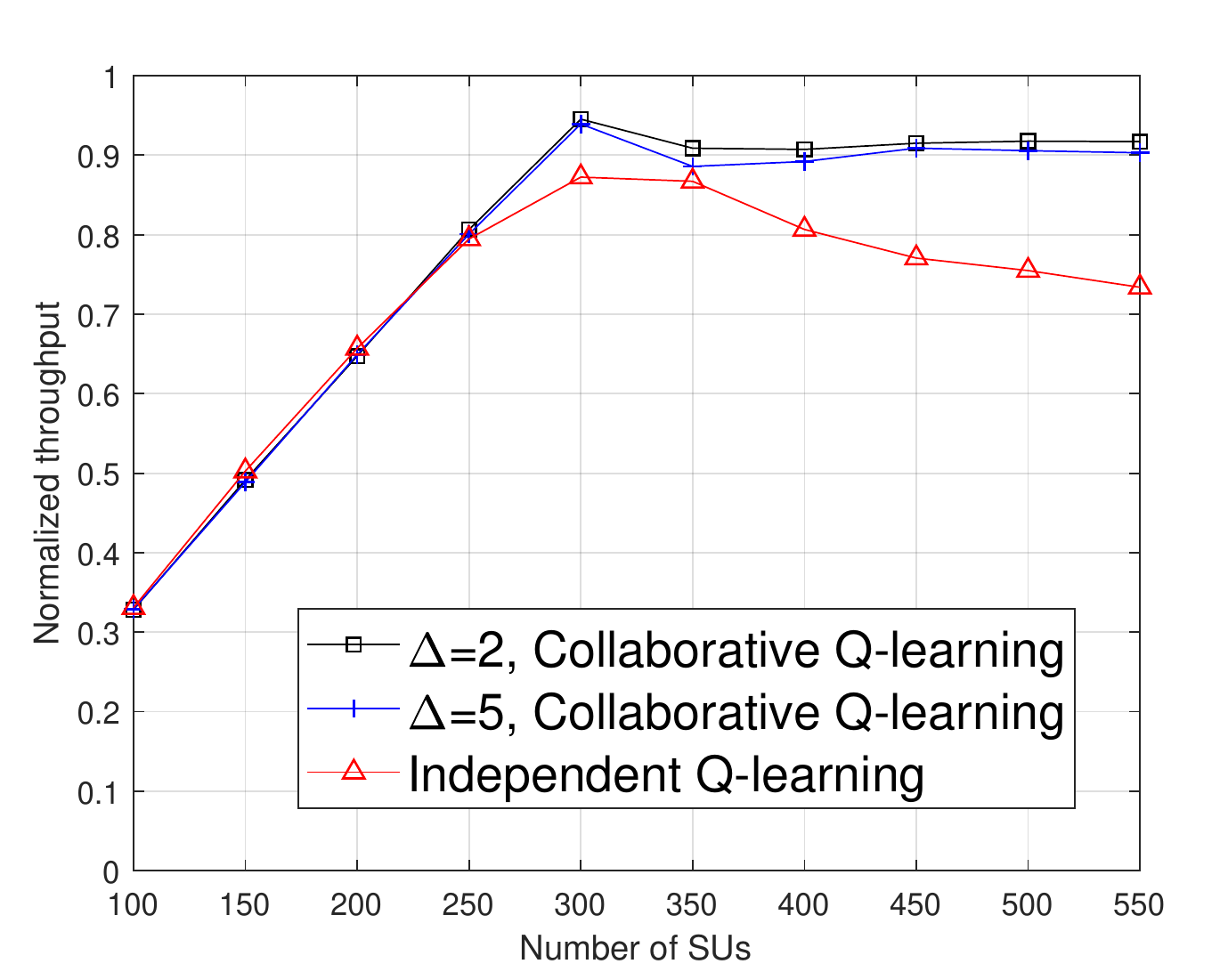}
\caption{Normalized throughput versus the number of SUs with the independent Q-learning and collaborative Q-learning-based schemes ($M = 300$, $\text{iterations} = 50$).}
\label{fig.1}
\end{figure}

\begin{figure}[!t]
\centering
\includegraphics[width=2.2 in]{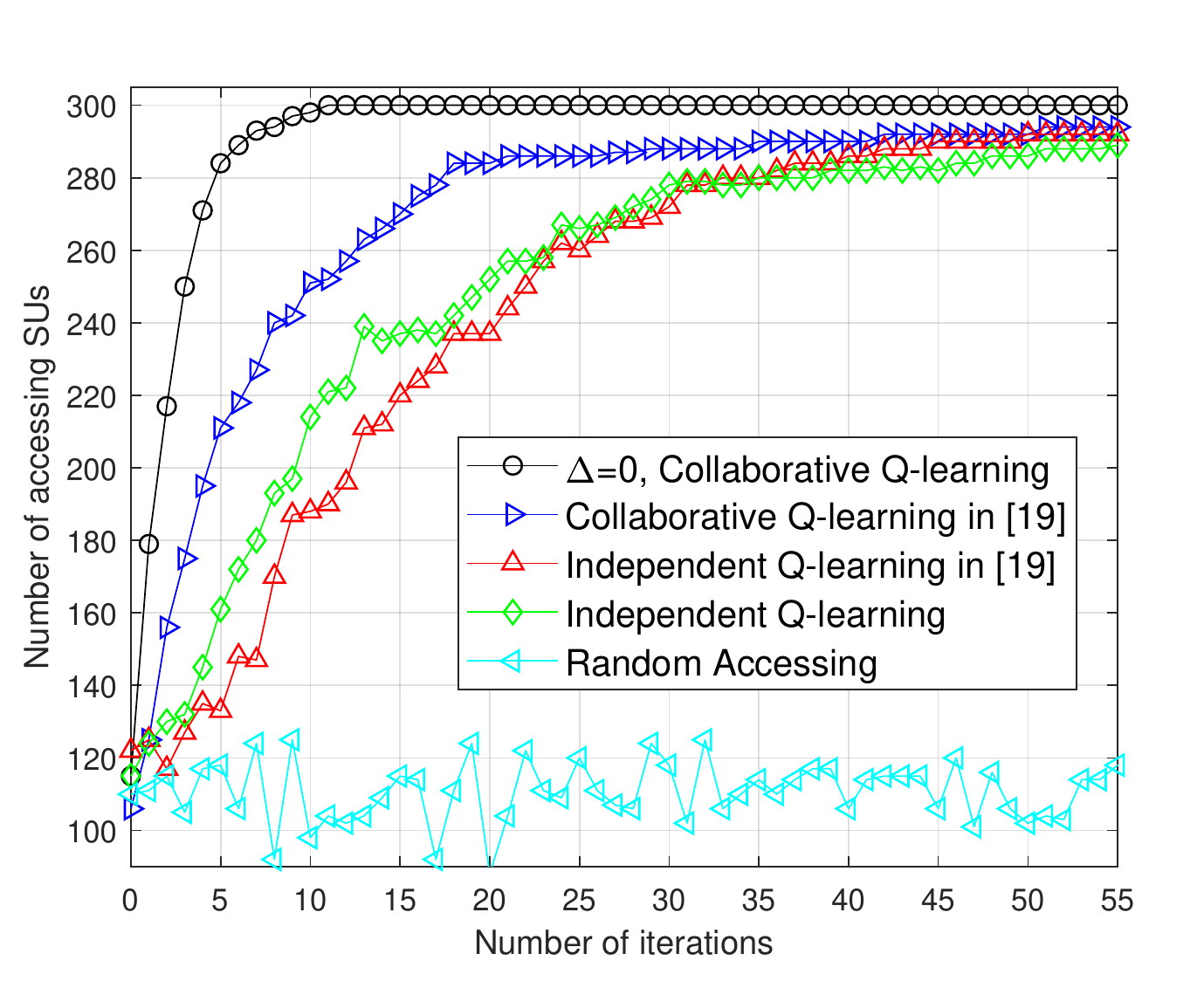}
\caption{The probabilities of SU and PU all blocking versus the numbers of SUs with the independent Q-learning and collaborative Q-learning-based schemes ($M = 300$, $\text{iterations} = 50$).}
\label{fig.1}
\end{figure}

\section{Simulation Results}

In this section, simulation results are presented to evaluate the performance of our proposed Q-learning-based subcarrier assignment schemes. A performance comparison between the independent Q-learning-based subcarrier assignment scheme and the distributed collaborative Q-learning with different information exchange intervals is presented. It is assumed that all involved channels are Rayleigh flat fading.

Our simulation parameters are set based on the work in \cite{Fuhui Zhou}, given as: $\bar{z}_{m,k}=0.1$, $\bar{g}_{m,k}=0.1$, $P_{m,k} = 10$ dB, $P_{m}^{PU}=10$ dB, $\alpha=0.2$ and the variance of additive white Gaussian noise $\sigma_{k}^{2}$ is 1. $\epsilon=0.8$ when $M>K$, otherwise, $\epsilon=0.1$. The number of subchannels, $M = 300$, which are competitively occupied by SUs at the beginning of the each frame. The number of iterations is 50 and the number of SUs is set within the interval [100, 550].

Fig. 2 shows the number of the successfully accessing SUs versus the number of iterations with the independent Q-learning and collaborative Q-learning in $\Delta=0$, $\Delta=2$ and $\Delta=4$. The number of PUs and that of SUs are both set as $300$. Fig. 2 shows the convergence of our proposed Q-learning schemes, and it is seen that the convergence of the collaborative Q-learning is better than that of the independent Q-learning scheme. The reason is the existence of information exchange among SUs. Specifically, when the available channels are not enough for SUs to access, most of the Q-values of SUs that exploit the collaborative Q-learning is less than zero. Therefore, they choose few channels that the Q-values are equal to zero. However, the performance of the SUs that exploit the independent Q-learning is limited by the inefficient exploration, so the exploration can cause interference to other channels. Moreover, it is seen that a faster convergence can be obtained with a smaller $\Delta$. However, a smaller $\Delta$ can result in a high complexity to update the Q-values. Thus, there exists a tradeoff between $\Delta$ and the computational cost.

Fig. 3 shows the probabilities of the SU and PU all blocking versus the numbers of SUs achieved with the independent Q-learning and collaborative Q-learning subcarrier assignment schemes. It is seen that, with the increase of the number of SUs, the probabilities of blocking of independent Q-learning increase rapidly when the number of SUs is greater than 300. However, for collaborative Q-learning that no mater $\Delta \text{ is } 2$ or $5$, the probabilities of blocking are still small.

Fig. 4 shows the normalized throughput versus the number of SUs with independent Q-learning and collaborative Q-learning subcarrier assignment schemes. Note that both collaborative and independent Q-learning schemes improve the throughput when the number of SUs is less than the number of subchannels. However, when $K>M$, the throughput achieved with the independent Q-learning method decreases rapidly. The normalized throughput achieved under the collaborative distributed Q-learning-based subcarrier assignment scheme does not decrease fast as that obtained under the independent Q-learning-based scheme. Another observation is that the normalized throughput achieved under the case of $\Delta=2$ is higher than that obtained under $\Delta=5$. The reason is that a lower information exchange frequency can result in that more SUs contend for the same subcarrier.

Fig. 5 shows the number of the successfully accessing SUs versus the number of iterations achieved with our proposed two Q-learning based methods and that obtained with the collaborative and independent Q-learning methods in \cite{S. K. Sharma}. The number of PUs and that of SUs are both set as 300 and the learning rates are all set as 0.1. It is seen that the convergence of the collaborative Q-learning with $\Delta=1$ is the best. The reason is that the collaborative Q-learning proposed in our work makes full use of the exchanging information. Another observation is that the convergent performance of the collaborative Q-learning in \cite{S. K. Sharma} is better than that of the independent Q-learning methods. The reason is that the collaborative method in \cite{S. K. Sharma} exchanges the congestion level, thus the agents know more about other agents than that with the independent Q-learning-based schemes. Moreover, the convergence of the independent Q-learning schemes is similar since they both know little about the environment.

\section{Conclusion}
In this paper, the subcarrier assignment problem was studied in wideband CRNs. An independent Q-learning-based subcarrier assignment and a distributed collaborative Q-learning scheme were proposed. Simulation results shown that there exists a tradeoff between the the computation cost and the information exchange interval. Moreover, it was shown that the performance achieved with the collaborative Q-learning-based method is better than that obtained with the independent Q-learning method.

\end{document}